# Recent claims on trapping antihydrogen are premature, if not false

G. Van Hooydonk, Ghent University, Faculty of Sciences, Ghent, Belgium

Abstract. We prove why claims on trapping antihydrogen are premature, if not false.

Recent claims by Enomoto et al. [1] and Andresen et al. [2] on trapping antihydrogen $\underline{H}$ are based on the hypothesis that $\underline{H}$ can be synthesized with long-range formation reaction

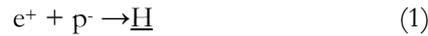

$$e^+ + p^- \rightarrow \underline{H} \qquad (1)$$

also the sole basis of all earlier H-claims [3-5]. Just by symmetry, (1) seems plausible by virtue of observed long-range charge-inverted reaction

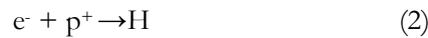

$$e^- + p^+ \rightarrow H \qquad (2)$$

for producing natural atom H. We give reasons why to falsify all $\underline{H}$-claims [1-5], based on (1).

(a) Claims on the synthesis of a novel species (atom, molecule) can only be validated if the species is identified with its spectrum. Theorists expect that H- and $\underline{H}$-spectra are identical. Any difference will be small, since Coulomb attraction $-e^2/r$ is identical in H and $\underline{H}$. The ultimate goal of [1-5] is to measure the $\underline{H}$-spectrum, e.g. its 1S-2S interval. Until hard spectral evidence is presented by which $\underline{H}$ can be probed, it is even impossible to claim, as in [1-5], that any $\underline{H}$ was actually produced. In the absence of the $\underline{H}$-spectrum, all $\underline{H}$-claims [1-5], based only on indirect evidence (annihilation-signatures), are premature.

(b) The second argument is that, if $\underline{H}$ were really mirrored H, spectral differences between left- and right-handed hydrogen are easily predicted, given the long history of left-right asymmetry, started in the 19th century (Pasteur, Le Bell, Van 't Hoff…). About 80 years ago, Hund [6] proved that these left-right differences follow a double well potential, i.e. a Mexican hat curve, whereby one well typifies the left-, the other the right-handed form of the same neutral species with identical formula and weight. Then, it is remarkable that the H Lyman series, available for many a decade, gives away this Mexican hat curve, typical for left-right behavior [6-8], see Fig. 1. This spectral evidence for left- and right-handed hydrogen states, available for long, was not considered by [1-5] before starting their experiments, based on (1).

(c) Correspondingly, the Hund-type Mexican hat curve in Fig. 1 not only proves the existence of H- and $\underline{H}$-states in natural atom hydrogen but it also reveals that the important interval 1S-2S belongs to the $\underline{H}$-domain. If so, the enigma about H and $\underline{H}$ spectra in (a) is avoided, although this is not realized in [1-5]. H- and $\underline{H}$-states separate at principal quantum number $n=\pi$ [7]; the height of the barrier is 0,05 cm$^{-1}$ (or 1,5 MHz). Since hydrogen is so abundant in the Universe, H- and $\underline{H}$-states are stable and annihilation must not occur for these neutral matter species (annihilation in the Dirac sense applies to charge-conjugated pairs with the same mass). Mass asymmetry avoids annihilation (see positronium as an example).

(d) The curve in Fig. 1 derives from monitoring the Rydberg $R_H$ or the ground state energy of hydrogen

$$R_H = -E_1 = -½e^2/r_1 \qquad (3)$$

in the complete interval between $n=1$ and $n=\infty$ (or $1/n=1$ and $1/n=0$). In fact, these n static ground state structures, with magnitude $n^2 E_n = -½e^2/r_1$ in Bohr theory [X], would give a fairly constant ratio

$$n^2E_n/E_1 = n^2E_n/(-\tfrac{1}{2}e^2/r_1) \approx 1 \qquad (4)$$

equal to about 1, provided there are no singularities in attractive lepton-nucleon Coulomb interaction – $e^2/r$, typifying both H, according to observed (2) and H̲, according to hypothetical (1), used in [1-5]. Fig. 2 gives the observed result for (4) in the complete interval $1/n=1$ to $1/n=0$. Using the ground state or series limit configuration $-E_1 = /(-\tfrac{1}{2}e^2/r_1) = 109678{,}7737$ cm$^{-1}$ as scaling factor, the observed Lyman terms learn that the observed Coulomb attraction of the lepton-nucleon pair fluctuates smoothly around its expectation value +1, with a maximum at $n=1{,}527226 \approx \tfrac{1}{2}\pi$ [X]. Although invisible in Bohr's achiral hydrogen theory, which cannot differentiate between Coulomb terms $-e^2/r$ for lepton-nucleon configurations (e⁻e⁺) and for (e⁺e⁻), these fluctuations indicate relatively small but essentially algebraic differences in the respective Coulomb attractions for H and H̲ of type $1/(1\pm x)$, with $x \ll 1$.

(d) Since the two curves in Fig. 1-2 expose the full lepton-nucleon reaction path for $\infty > n > 1$, observed dissociation products e⁻, p⁺ prove that their interaction is confined to long range $\infty > n > \pi$ to give H (well at –1 in Fig. 1 or $1/(1+x)<1$ in Fig. 2). It equally follows that (1) is confined to short range $\pi > n > 1$ to give H̲ (well at +1 in Fig. 1 or $1/(1-x)>1$ in Fig. 2). The resulting conclusive schema is shown in Fig. 3

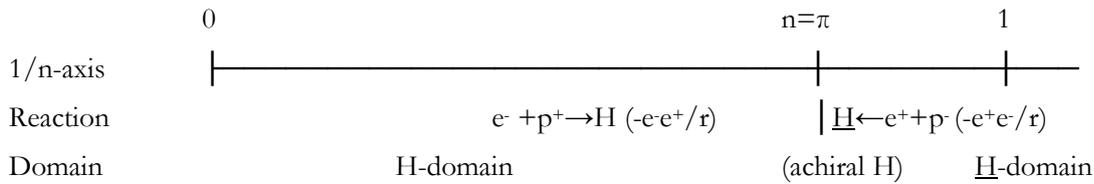

Fig. 3 Confined n-domains for H, reaction (2) and for H̲, reaction (1), separated by achiral hydrogen.

Then, long-range reaction (1), used in [1-5] to synthesize H̲, is forbidden, if not impossible in nature [9]. In fact, it would reverse the natural order of two states (say two phases of aggregation). This does not make sense, as it would imply for example that ice could be made by heating water [9].

Hence, claims [1-5] are not only premature; they are probably also false, as argued prior [7-10] to the first 2002 H̲-claims by ATHENA and ATRAP [3].

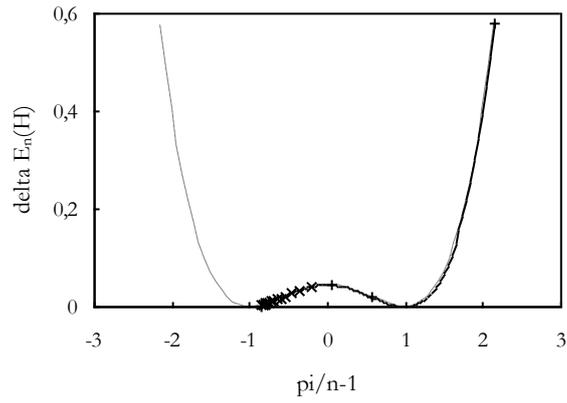

Fig. 1. H Lyman series double well or Mexican hat curve $\delta=R/n^2-E_n$ ($R=109679,35$ cm$^{-1}$ [8]) versus $\pi/n-1$
($n<\pi$: antiH-states x, $n>\pi$: H-states +)), adapted from [7]

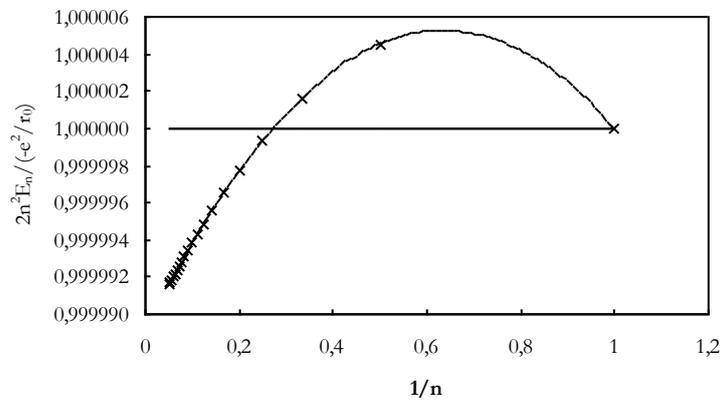

Fig. 2 Observed $n^2E_n/E_1\approx1$ in (4) between $1/n=0$ and $1/n=1$ [8] (for comparison, a full line is drawn at expectation value +1)